\documentclass[aps,prl,twocolumn,showpacs,superscriptaddress,groupedaddress]{revtex4} 
\usepackage[latin1]{inputenc}
\usepackage{graphicx}
\usepackage{amssymb}
\usepackage{color}
\usepackage{float}
\usepackage{amsmath}
\usepackage{amsfonts}
\usepackage{dcolumn}
\usepackage{hyperref}
\usepackage{amsthm}
\usepackage{color}

\def\3nab{\tilde{\nabla}}

\def\be {\begin{equation}}
\def\ee {\end{equation}}
\def\ba {\begin{eqnarray}}
\def\ea {\end{eqnarray}}

\newcommand{\bra}[1]{\left(#1\right)}

\newcommand{\E}{{\mathcal E}}

\newcommand{\barray}{\begin{array}}
\newcommand{\earray}{\end{array}}

\newcommand{\udot}{{\mathcal A}}

\begin{document}
 
\title{Thermodynamics of gravity favours Weak Censorship Conjecture}

\author{Giovanni Acquaviva}
 \email{acquavivag@unizulu.ac.za}
 \affiliation{Department of Mathematical Sciences, University of Zululand, Private Bag X1001, Kwa-Dlangezwa 3886, South Africa.}
\author{Rituparno Goswami}
 \email{goswami@ukzn.ac.za}
\author{Aymen I. M. Hamid}
 \altaffiliation{Physics Department, University of Khartoum, Sudan.}
 \email{aymanimh@gmail.com}
 \affiliation{Astrophysics and Cosmology Research Unit, School of Mathematics, Statistics and Computer Science, University of KwaZulu-Natal, Private Bag X54001, Durban 4000, South Africa.}
 \author{Sunil D. Maharaj}
\email{Maharaj@ukzn.ac.za}
 \affiliation{Astrophysics and Cosmology Research Unit, School of Mathematics, Statistics and Computer Science, University of KwaZulu-Natal, Private Bag X54001, Durban 4000, South Africa.}

\begin{abstract}
We use the formulation of thermodynamics of gravity as proposed by Clifton, Ellis and Tavakol on the gravitational collapse of dustlike matter, that violates the strong or weak cosmic censorship conjecture depending on the initial data. We transparently demonstrate that the gravitational entropy prefers the scenario where the stronger version is violated but the weak censorship conjecture is satisfied. This is a novel result, showing the weak cosmic censorship and hence the future asymptotically simple structure of spacetime, is being validated by the nature of gravity, without imposing any extra constraint on the form of matter.
 
\end{abstract}

\pacs{04.20.Cv, 04.20.Dw}

\maketitle


Since the inception of Cosmic Censorship Conjecture (CCC) more than four decades ago \cite{CCC}, there have been numerous unsuccessful attempts towards its precise mathematical proof, within the context of general relativity \cite{wald, Joshibook1,Joshibook2}. The reason for the continuing endeavour for establishing CCC on a solid mathematical footing is obvious; this conjecture is one of the key ingredients for the proofs of all important theorems of black hole dynamics and thermodynamics \cite{HE}. Therefore any transgression of CCC would imply a reformulation of otherwise well established results on the structure of black hole physics, that has led to many remarkable theoretical developments. Hence comes the necessity of the theoretical censoring of {\em naked singularities} in nature.

As of now, there are two versions of CCC that are used extensively in the literature \cite{wald}. The physical formulation of the weaker one (WCC) states that {\it ``All singularities of a gravitational collapse are hidden within black holes and cannot be seen by a distant observer''}. The stronger version (SCC) is physically formulated as follows: {\it ``All physically reasonably spacetimes are globally hyperbolic. That is, apart from possible initial singularities like Big Bang, no singularity is ever visible to any observer''}. In the context of gravitational collapse of astrophysical compact stars, the impact of the above conjectures can be summarised in the following way. The stronger version asserts that the spacetime singularity thus formed must be immersed in the trapped region and hence there cannot be any outgoing future directed non-spacelike geodesics from the vicinity of the singularity. Whereas the weaker version admits the possibility of the existence of such geodesics, but these will definitely enter the trapped region in future and fall back to the singularity, thus saving the future asymptotic structure of the spacetime manifold.

In spite of the absence of any mathematical proof of cosmic censorship, numerous studies have produced a large number of counterexamples, where one or both of the above versions are violated during dynamical gravitational collapse of different forms of matter fields under very generic conditions (see \cite{Goswami:2006ph, Joshibook1,Joshibook2, Mkenyeleye:2014dwa} for detailed references). The main reason for violation of CCC in these counterexamples is that the trapped surfaces do not form early enough to shield the singularity from outside observers. One possible way to to avoid the unpleasantries of cosmic nudity would be to assert that the form of matter fields, that lead to the violation of censorship, are {\em unphysical}. However, these counterexamples span a wide class of matter fields (namely perfect fluids, null fluids, scalar fields or a combination of these) obeying all physically reasonable energy conditions and such a strong restriction on the thermodynamics of matter fields would itself be unphysical. Furthermore, in a recent investigation \cite{Hamid:2014kza}, it was shown that the Weyl curvature (which is a measure of {\em free} gravity) plays an important role in the violation of censorship.

Under these circumstances the key question that arises here is as follows: {\em Instead of thermodynamics of matter, should the thermodynamics of the gravitational field dictate the final outcome (in terms of a black hole or a naked singularity) of the gravitational collapse?} Although this is a very important question, until recently it would have been impossible to investigate upon, due to the unavailability of suitable local definitions of thermodynamic quantities of the gravitational field.  In fact, Penrose \cite{Pen79,Pen:10} did suggest that the definition should be based in properties of the Weyl tensor (instead of the Ricci tensor which is directly related to matter fields via Einstein field equations) but gave no specific formula. Recently, in a seminal work \cite{Clifton:2013dha}, it was proposed that the energy density and pressures for the gravitational field can be extracted from the square root of the Bel-Robinson tensor \cite{Bel-Robinson}, which is constructed from the Weyl tensor and it's dual and acts like an effective energy-momentum tensor of the free gravitational field \cite{Maartens:1997fg}. It was also proposed that the local temperature of the gravitational field for an observer with 4-velocity $u^a$ can be generalised from the expression of the surface gravity \cite{Bardeen:1973gs}, which is given as
\be\label{temp}
T_{grav}=\frac1\pi l^ak^b\nabla_bu_a\;,
\ee
where $l^a$ and $k^a$ are real null vectors in a Newman-Penrose tetrad. These definitions have been shown to satisfy the necessary conditions in order to generate a well defined entropy and have already been applied in different scenarios: the analysis of scalar perturbations in FLRW spacetime \cite{Clifton:2013dha,larena}, the growth of inhomogeneities in perfect fluids, Oppenheimer-Snyder dust collapse and its final black hole stage \cite{Acquaviva:2014owa}.  The results obtained are either consistent with previous studies or in line with expected thermodynamic behaviour \cite{Bek73}.  In particular it has been possible to prove that the formation of structures under gravity is a thermodynamically favoured process and that the usual Bekenstein-Hawking entropy is the final result of the change in gravitational entropy during collapse that eventually forms the black hole.

Through these definitions, it was shown that it is possible to define a first law of thermodynamics for the gravitational field as
\be\label{2nd}
 T_{grav} \dot{S}_{grav} = \left(\mu_{grav}v\right)^.+p_{grav}\dot{v}\, ,
\ee
where $T_{grav}$, $\mu_{grav}$, $p_{grav}$ and $S_{grav}$ represent the effective temperature, energy density, isotropic pressure and entropy density of the gravitational field.  Here $v$ is the spatial volume element and the dot represents the derivative with respect to time.  For Petrov type D spacetimes we have $p_{grav}=0$ and the other thermodynamic quantities \cite{Bardeen:1973gs} can be written as
\begin{equation}\label{dels}
 dS_{grav}=\frac{d\bra{\mu_{grav} v}}{T_{grav}},\ \ \mu_{grav}=\frac{1}{8\pi} |\E|,
\end{equation}
 where $\E$ is the Weyl curvature scalar constructed via the semi-tetrad 1+1+2 covariant formalism \cite{Clarkson:2002jz,Clarkson:2007yp,Betschart:2004uu}.    Furthermore, if we assume spherical symmetry, in the semi-tetrad formalism the expression of the gravitational temperature (\ref{temp}) becomes
\be
T_{grav}=\frac{|\udot+\frac{1}{3}\Theta+\Sigma|}{2\pi}\;,
\label{tem1+1}
\ee
where $\udot$ is the 4-acceleration, $\Theta$ is the volume expansion and $\Sigma$ is the shear scalar of the congruence of the observer's 4-velocity $u^a$.
This quantity can be identified as the temperature of the gravitational field measured locally at a point of the spacetime, associated with the 2-sphere through that point.
Once all quantities are specified, one can integrate eq.\eqref{dels} on a spacelike hypersurface to obtain the instantaneous measure of gravitational entropy $S_{grav}$ for that specific time slice.  In the following we will be concerned also with the variation of gravitational entropy between two different configurations $\delta S_{grav} = S_{grav}(t_2)-S_{grav}(t_1)$, where $t$ is the relevant timelike coordinate (an affine parameter to the integral curves of $u^a$).

Now, to clearly see whether the thermodynamics of free gravity indeed have a preference towards CCC or otherwise, we need an exact model of  a collapsing star that obeys/disobeys either version of censorship conjecture, depending on the initial data at the time slice from which the collapse commences. The simplest example is a spherically symmetric collapsing star of marginally bound dustlike matter, that can be smoothly matched to a vacuum Schwarzschild exterior. In this case, the metric in the interior of the star will be the marginally bound Lemaitre-Tolman-Bondi metric in comoving $(t,r,\theta,\phi)$ coordinates:
 \be\label{ltb}
ds^2=-dt^2+R'(t,r)^2dr^2+R(t,r)^2\left( d\theta^2+\sin^2\theta d\phi^2\right),
\ee
where  $R(t,r)$ is the area radius of the collapsing spherical dust shell labeled by $r$ at an instant $t$ and the prime denotes derivative with respect to $r$. By the scaling freedom of the radial coordinate, we can always scale the coordinate $r$ such that at initial epoch $t=t_i$, from which the collapse commences, we have $R(t_i,r)=r$. This collapsing star is smoothly matched to the Schwarzschild exterior at the boundary shell denoted by $r=r_b$.
From the $(t,t)$ component of Einstein field equations for the metric \eqref{ltb} we get $F'=\rho R^2R'$, where $F=F(r)$ is the Misner-Sharp mass \cite{Misner:1964je} of the star which is related to the Schwarzschild mass $M$ of the exterior spacetime by $F(r_b)=2M$, and $\rho(t,r)$ is the density of the collapsing star. The mass function can be related to the initial density $\rho_0(r)\equiv\rho(t_i,r)$ by the relation $F(r)=\int\limits_0^r x^2\rho_0(x)dx$. Thus we can always write $F(r)\equiv r^3{\cal M}(r)$, where ${\cal M}(r)$ is a well defined function $\forall r\in[0,r_b]$.

For the marginally bound case, the equation of motion for the collapsing shells is given by $R\dot{R}^2=F$ \cite{Goswami:2004gy, Goswami:2002he}.
Solving the equation of motion we obtain
\be\label{a}
R(r,t)=r\, \bra{1-\frac{3}{2}\sqrt{{\cal M}(r)}t}^{2/3}.
\ee
To determine the time at which every shell crosses the horizon and becomes trapped, from the equation of the apparent horizon $F=R$, we extract the general expression
\be\label{tah}
 t_{ah}(r) = \frac{2}{3 \sqrt{{\cal M}(r)}}-\frac{2}{3}\, F(r)\, .
\ee
To demonstrate transparently the role of initial data in the violation of CCC, let us consider a realistic initial density profile: $\rho_0(r)=\rho_{00}-\rho_{02}r^2$. To avoid pathologies like shell crossing singularities, and to mimic realistic stars where the density is a non-increasing function of the radial coordinate, we must have $\rho_{02}\ge0$. This initial data will then imply ${\cal M}(r)=(\rho_{00}/3)-(\rho_{02}/5)r^2\equiv m_0-m_2\, r^2$. A central singularity forms at $t_{s_0}=2/(3\sqrt{m_0})$ while the outermost shell $r_b$ collapses in the singularity at time $2/(3\sqrt{{\cal M}(r_b)})$.

\begin{figure}[htb!!!]
 \begin{center}
 \includegraphics[width=4.25cm]{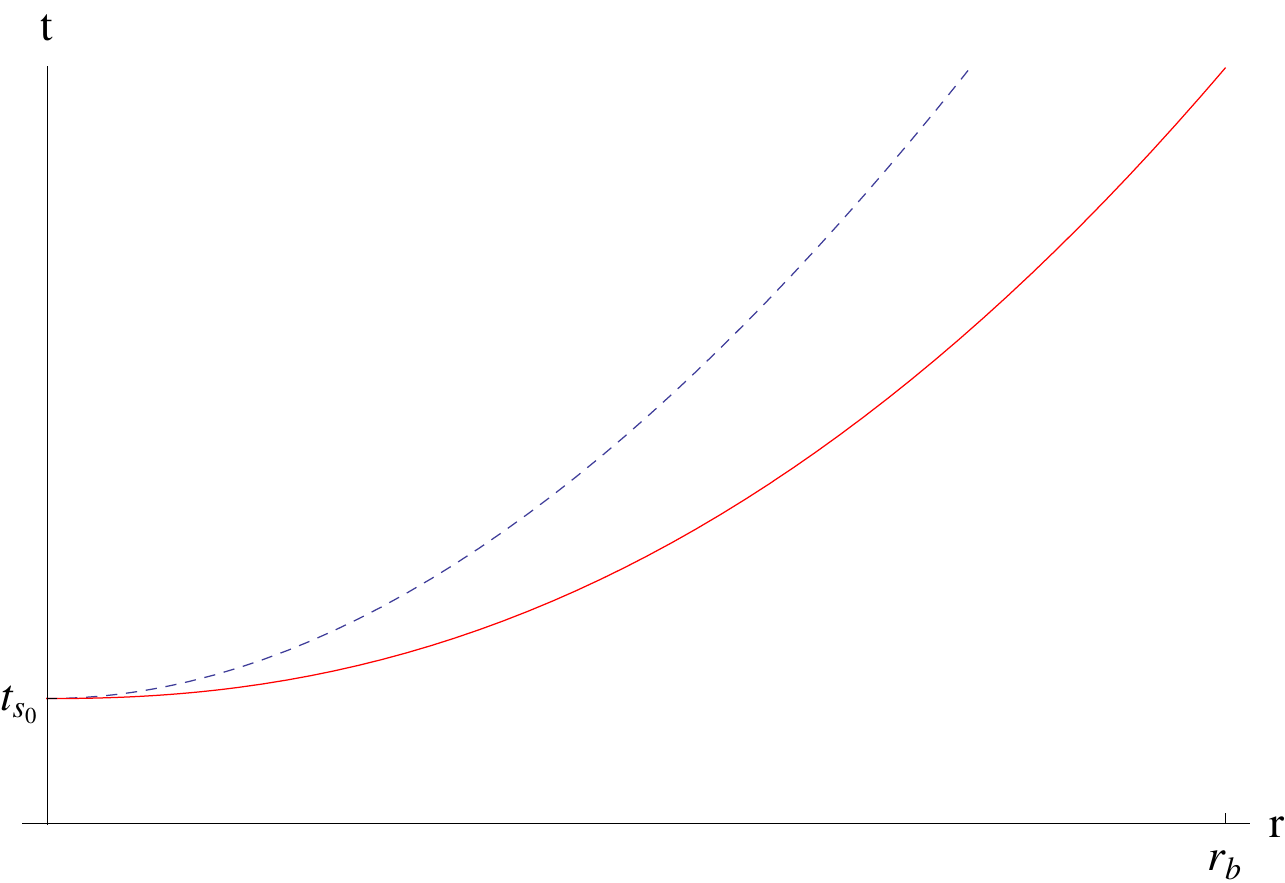}
 \includegraphics[width=4.25cm]{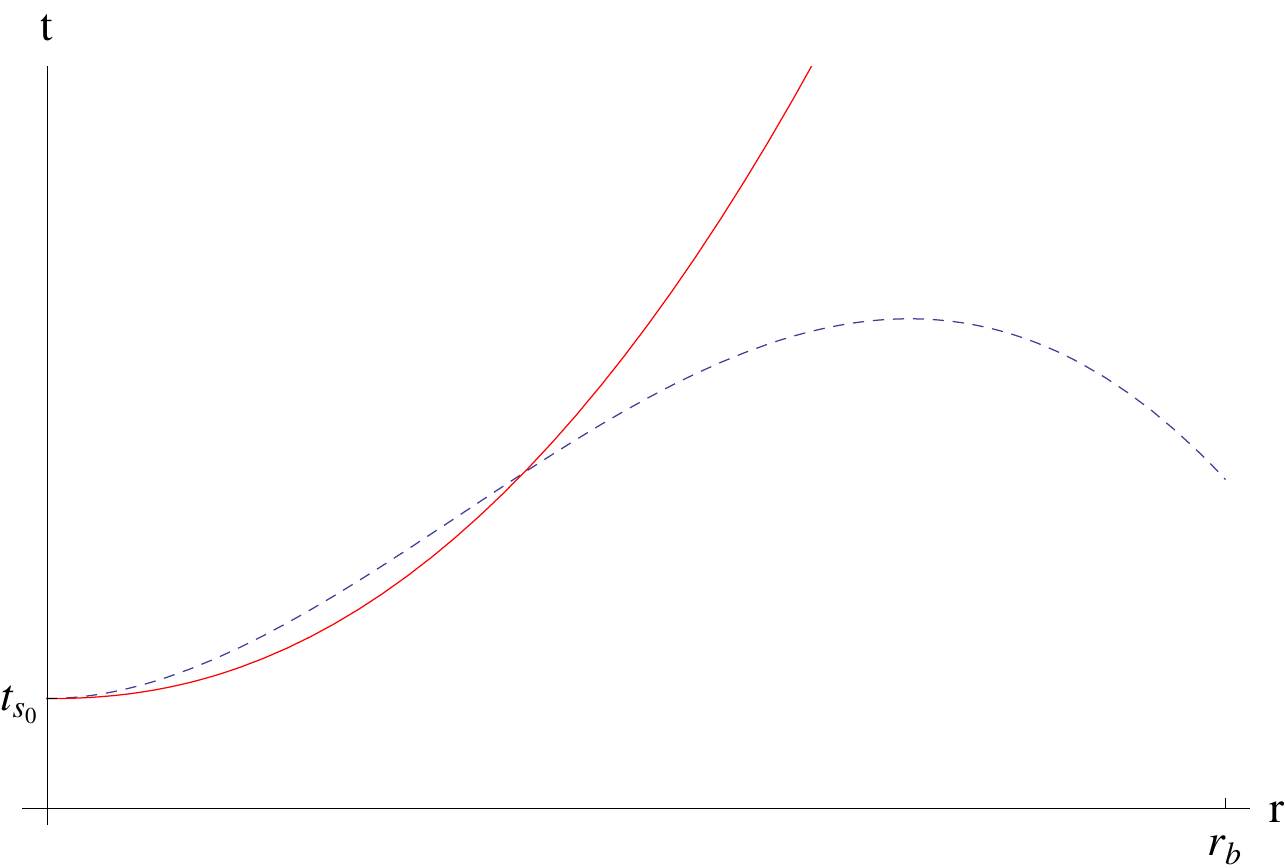}
 \caption{The two scenarios depicting globally naked (left) and locally naked (right) situations. The red solid line is the outgoing null geodesic and the blue dashed line is the apparent horizon in $(t,r)$ plane.}
\end{center}
\end{figure}

It has been shown quite rigorously by many authors (see for example \cite{Joshibook2,Goswami:2002he,Goswami:2006ph} and the references therein) that whenever $\rho_{02}>0$ the central singularity is locally naked with a future directed radial null geodesic emerging from it. The equation for this null geodesic $t_{ng}(r)$ in the $(t,r)$ plane can be obtained by integrating the null geodesic equation $(dt/dr)_{ng}=R'$, with the initial value arbitrarily close to the central singularity. The existence of such null geodesic implies violation of SCC. However WCC may still be satisfied if the null geodesic enters the trapped region in the future at $r=r^*$, where $r^*$ is the smallest real positive root of the equation $t_{ng}(r)=t_{ah}(r)$. Thus, with the given initial data, we may have three distinct final outcomes of the collapse:
\begin{enumerate}
\item {\bf Case I:} {{$\rho_{02}=0$}}. This is the well known case of Oppenheimer-Snyder-Datt \cite{osd,datt} collapse and the vicinity of the central shell gets trapped before the singularity formation. In this case, there cannot be any outgoing null geodesics from the central singularity and hence both SCC and WCC are satisfied.
\item {\bf Case II:} {{$\rho_{02}>0, r_b>r^*$}}.  In this case the central singularity is locally naked, however the future directed null geodesic from the singularity enters the trapped region sometime in the future and falls back to the singularity. In this case SCC is violated but WCC is still satisfied and the asymptotic structure of the spacetime does not change.
\item{\bf Case III:} {$\rho_{02}>0, r_b<r^*$}.  In this case the null geodesic from the naked central singularity leaves the stellar surface before being trapped and travels to future null infinity, destroying the asymptotic structure. In this case both SCC and WCC are violated and the central singularity is globally naked. Figure 1 shows the difference between a locally and globally naked singularity.
\end{enumerate}

To examine which of these three cases are most favoured by the thermodynamics of gravity, we calculate the net entropy change from the initial epoch to the epoch of the central singularity $t=t_{s_0}$, as measured by congruences of free falling observers both in the interior of the star and the exterior Schwarzschild spacetime. To do this, we must exactly match the time slices of interior and exterior spacetimes, which can be done by describing the Schwarzschild geometry in Lema\^{i}tre coordinates
 \be\label{sch}
ds^2=-dt^2+\frac{2M}{\tilde{R}}\, d\tilde{r}^2+\tilde{R}^2\ d\Omega^2\, ,
\ee
where the area radius is given by $\tilde{R}(\tilde{r},t)=\left( \frac{3}{2}(\tilde{r}-t)\right)^{2/3}$, and the comoving radial coordinate $\tilde{r}$ labels each free falling trajectory.

In order to obtain the gravitational entropy for the exterior spacetime, we need to integrate eq.\eqref{dels} on a spacelike hypersurface, with gravitational energy density and temperature given respectively by
\be
\mu_{grav}=\frac{1}{8\pi}\, \frac{2M}{\tilde{R}},\;T_{grav}=\frac{1}{4\pi}\, \frac{\sqrt{2M}}{\tilde{R}^{3/2}},
\ee
where we have removed the absolute values since $\tilde{R}$ and $M$ are both positive.  Therefore the total gravitational entropy over a generic time slice of the exterior spacetime reduces to
\be\label{deltasin}
S^{ext}_{grav}(t)=4\pi M\left(\tilde{r}_{\infty}-\tilde{r}_s\right)\\, 
\ee
where $\tilde{r}_s$ is the coordinate of the free falling collapsing stellar surface, while $\tilde{r}_{\infty}$ denotes the free falling observer at infinity. 
This implies that the change in gravitational entropy in the exterior spacetime between any two comoving time slice as measured by congruences of free falling observers vanishes:
\be
\delta S^{ext}_{grav}=S^{ext}_{grav}(t_2)-S^{ext}_{grav}(t_1)\equiv0.
\ee
Now for the interior spacetime we can easily calculate the relevant quantities in eq.\eqref{dels} and eq.\eqref{tem1+1} to get
\begin{align}
\mu_{grav}&=\frac{1}{8 \pi}\left|\frac{1}{3} \frac{F'}{R'\, R^2}-\frac{F}{R^3}\right|\;,\\
T_{grav}&=\frac{1}{2\pi}\left|\frac{\dot{R}'}{R'}\right|\,.
\end{align}
After some manipulations, we can write the gravitational entropy at any time slice ``$t$" in the following useful form:
\begin{equation}\label{sgltb}
 S^{int}_{grav}(t)=2\pi\int\limits_0^{r_b}\left|\frac{{\cal E}/\rho}{{\cal E}/\rho+\frac{2}{3}}\right|\, R'\, \sqrt{F\, R}\ dr\,.
\end{equation}
Figure 2 shows the numerically integrated results of $\delta S^{int}_{grav}\equiv S^{int}_{grav}(t_{s_0})-S^{int}_{grav}(t_i)$ as a function of the boundary of the star $r_b$ for three pairs of $m_0$ and $m_2$.  We recall that when $r^*=r_b$ the singularity is {\em marginally locally naked}.  It can be easily seen that the change in total entropy is a monotonically increasing function of the boundary radius of the collapsing star, for physically realistic values of the initial density parameters.
 \begin{figure}[htb!!!]
 \begin{center}
 \includegraphics[width=8cm]{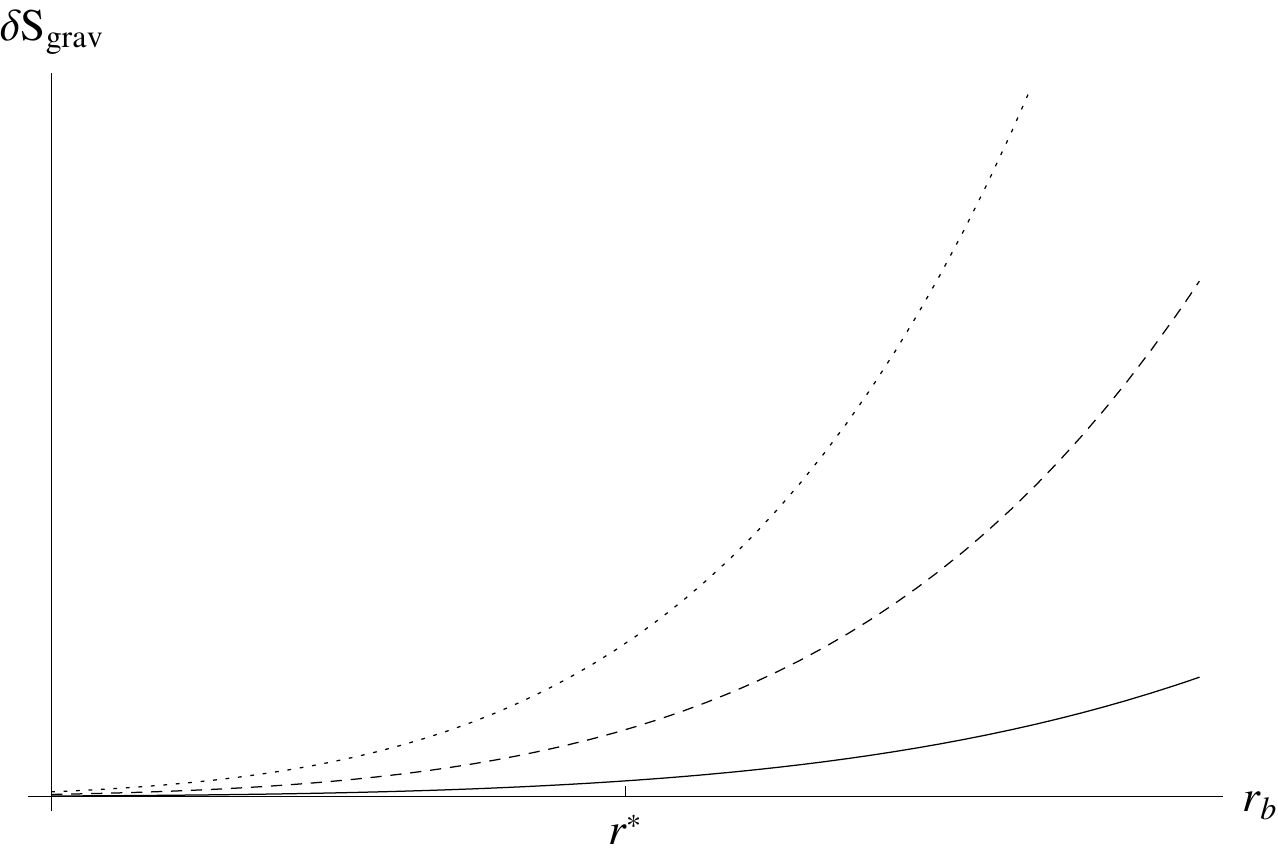}
 \caption{\label{deltas} Numerical integration of $\delta S^e_{grav}(r_b)$ in a neighbourhood of $r^*=0.2$ for different points $\{m_0,m_2\}$ in the parameter space: $\{0.3,0.05\}$ (dotted), $\{0.2,0.018\}$ (dashed) and $\{0.1,0.003\}$ (solid).}
\end{center}
\end{figure}

Now, let us revisit the three distinct cases mentioned before, in light of the net gravitational entropy change during the process of collapse. We would like to emphasise that since dust is an isentropic form of matter, there will be no change in matter entropy for the collapsing star. Hence, change in the gravitational entropy will account for the total entropic evolution. For the first case, where both SCC and WCC are satisfied, since the Weyl tensor in the interior spacetime is identically zero, there is no change in gravitational entropy, and hence this will be the least favoured case. Now to compare between the second and third cases, we can see from Figure 2 that the net gravitational entropy change for $r_b>r^*$ is greater than that for $r_b<r^*$. Therefore the second case, in which SCC is violated but WCC is not, will in general be more favoured thermodynamically.

This is indeed a very interesting result, as it clearly shows that the thermodynamics of gravity discards the homogeneous case which is clearly unstable under matter perturbations.  Moreover, it is fascinating to see that gravity is favouring the scenario that saves the asymptotic structure of the spacetime, validating most of the important theorems of black hole dynamics and thermodynamics. This result leads us to argue that any general mathematical proof of censorship conjectures should naturally emerge from the thermodynamic properties of the gravitational field, a detailed study of which will be reported later.

\begin{acknowledgments}
GA is thankful to the Astrophysics and Cosmology Research Unit (ACRU) at the University of KwaZulu-Natal for the kind hospitality.  AH and  RG  are supported by National Research Foundation (NRF), South Africa. SDM acknowledges that this work is based upon 
research supported by the South African Research Chair Initiative of the Department of Science and Technology.
\end{acknowledgments}


\begin{thebibliography}{99}

\bibitem{CCC}
R. Penrose, 
Riv. Nuovo Cimento, Num. Sp. I, 1969.


\bibitem{wald} R. M. Wald, {\em General Relativity}, (University of Chicago Press, Chicago, 1984).

\bibitem{Joshibook1} 
P.  S. Joshi, {\it Global Aspects in Gravitation and Cosmology} (Oxford University Press, 1993).



  
\bibitem{Joshibook2} 
 P. S. Joshi,{ \it Gravitational Collapse and Spacetime Singularities} (Cambridge University press, 2007).
 
\bibitem{HE}
S. W. Hawking and G. F. R.
Ellis, \emph{\it The Large Scale Structure of Spacetime}
(Cambridge University Press, 1973).

 

\bibitem{Goswami:2006ph}
  R.~Goswami and P.~S.~Joshi,
  Phys.\ Rev.\ D {\bf 76} (2007) 084026
  [gr-qc/0608136].

  
\bibitem{Mkenyeleye:2014dwa} 
  M.~D.~Mkenyeleye, R.~Goswami and S.~D.~Maharaj,
  Phys.\ Rev.\ D {\bf 90}, no. 6, 064034 (2014)
  [arXiv:1407.4309 [gr-qc]].



\bibitem{Hamid:2014kza}
  A.~I.~M.~Hamid, R.~Goswami and S.~D.~Maharaj,
  Class.\ Quant.\ Grav.\  {\bf 31} (2014) 135010
  [arXiv:1402.4355 [gr-qc]].
  


\bibitem{Pen79}
R. Penrose, "Singularities and Time-Asymmetry". In S. W. Hawking and W. Israel. "General Relativity: An Einstein Centenary Survey" (Cambridge University Press, 1979)



\bibitem{Pen:10} R. Penrose \textit{Cycles of Time: An Extraordinary New View of the Universe} (Bodley Head, London, 2010).

\bibitem{Clifton:2013dha} 
  T.~Clifton, G.~F.~R.~Ellis and R.~Tavakol,
  Class.\ Quant.\ Grav.\  {\bf 30}, 125009 (2013)
  [arXiv:1303.5612 [gr-qc]].
  
\bibitem{Bel-Robinson} 
  L. Bel, C.R. Acad Sci. Paris 247, 1094-1096 (1958)
  
  
\bibitem{Maartens:1997fg} 
  R.~Maartens and B.~A.~Bassett,
  Class.\ Quant.\ Grav.\  {\bf 15}, 705 (1998)
  [gr-qc/9704059].

\bibitem{Bardeen:1973gs} 
  J.~M.~Bardeen, B.~Carter and S.~W.~Hawking,
  Commun.\ Math.\ Phys.\  {\bf 31}, 161 (1973).

 
 \bibitem{larena}
 R.~A.~Sussman and J. Larena,
Class. and Quantum Grav. 31.7 (2014): 075021.

  
 
\bibitem{Acquaviva:2014owa}
  G.~Acquaviva, G.~F.~R.~Ellis, R.~Goswami and A.~I.~M.~Hamid,
  Phys.\ Rev.\ D {\bf 91} (2015) 6,  064017
  [arXiv:1411.5708 [gr-qc]].


\bibitem{Bek73}
J.D. Bekenstein, 
Phys. Rev. D {\bf 7} (1973) 2333-2346. 

  
\bibitem{Clarkson:2002jz} 
  C.~A.~Clarkson and R.~K.~Barrett,
  Class.\ Quant.\ Grav.\  {\bf 20}, 3855 (2003)
  [gr-qc/0209051].
  
\bibitem{Betschart:2004uu} 
  G.~Betschart and C.~A.~Clarkson,
  Class.\ Quant.\ Grav.\  {\bf 21}, 5587 (2004)
  [gr-qc/0404116].
  
\bibitem{Clarkson:2007yp} 
  C.~Clarkson,
  Phys.\ Rev.\ D {\bf 76}, 104034 (2007)
  [arXiv:0708.1398 [gr-qc]].

\bibitem{Misner:1964je} 
  C.~W.~Misner and D.~H.~Sharp,
  Phys.\ Rev.\  {\bf 136}, B571 (1964).
  
\bibitem{Goswami:2002he} 
  R.~Goswami and P.~S.~Joshi,
  Phys.\ Rev.\ D {\bf 69}, 044002 (2004)
  [gr-qc/0212097].
  
 
\bibitem{Goswami:2004gy}
  R.~Goswami and P.~S.~Joshi,
  Phys.\ Rev.\ D {\bf 69} (2004) 104002
  [gr-qc/0405049].
  
 
 \bibitem{osd} 
J. R. Oppenheimer and H. Snyder, 
Phys. Rev. {\bf 56}:455 (1939).

  \bibitem{datt} 
  B. Datt 
Z. Phys. {\bf 108} (1938), reprinted as a Golden Oldie, General Relativity and Gravitation, 31, 1615 (1999).



%
%
%
%
%
%
%
%
%
%
%
%
%
%
%
%
%
%
%
%
%
%
%
%
%
%
%
%
%
%
%
%

  

\end{thebibliography}
\end{document}